\documentclass[preprint,aps,prd]{revtex4-1}
\usepackage{graphicx}
\usepackage{slashed}
\usepackage{verbatim} 
\newcommand{\be}{\begin{eqnarray}}
\newcommand{\ee}{\end{eqnarray}}
\usepackage[colorlinks]{hyperref}
\usepackage{cancel,bm} 
\usepackage{amsmath,amssymb}
\usepackage{mathtools}
\usepackage{float}
\usepackage{color}
\usepackage{leftidx}
\usepackage{booktabs}
\usepackage{latexsym}
\usepackage{revsymb}
\usepackage{multirow}
\usepackage{hypcap}
\usepackage{array}
\hypersetup{
    colorlinks=red,
    citecolor=blue,
    filecolor=magenta,      
    urlcolor=red,
}

\begin{document}

\title{Accessing Linearly Polarized Gluon Distribution  in $J/\psi$ Production at the Electron-Ion Collider}

\author{Raj Kishore and Asmita Mukherjee}
\affiliation{ Department of Physics,
Indian Institute of Technology Bombay, Mumbai-400076,
India.}
\date{\today}

\begin{abstract}
We calculate the $cos~2 \phi_h$ asymmetry in $J/\psi$ production in electron-proton collision for the kinematics of the planned electron-ion collider (EIC). This directly probes the Weisz\"acker-Williams (WW) type linearly polarized gluon distribution. Assuming generalized  factorization, we calculate the asymmetry  at next-to-leading-order (NLO) when the energy fraction of the $J/\psi$  satisfies $z<1$ 
and the dominating subprocess is $\gamma^* +g \rightarrow c + {\bar c}+g$.  We use non-relativistic QCD based color singlet (CS) model for $J/\psi$ production.  We investigate the  small $x$
region which will be accessible at the EIC.  We present the upper bound of the asymmetry, as well as  estimate it using a (i) Gaussian type parametrization for the TMDs  and (ii) McLerran-Venugopalan (MV) model at small $x$.  We find  small but sizable asymmetry in all the three cases.

\end{abstract}

\maketitle
\raggedbottom 

\section {Introduction}

$J/\psi$ electroproduction is a direct probe of gluon transverse momentum dependent
parton distributions (TMDs), as the leading process is the virtual photon-gluon
fusion. Very little is known so far about the gluons TMDs
\cite{Mulders:2000sh}, apart from a
positivity bound. Recently, unpolarized TMD gluon pdfs have been extracted
from LHCb data \cite{Lansberg:2017dzg}. TMD pdfs are process dependent due to 
the initial and/or final state interactions, in other words, due to the presence 
of gauge links in their operator definitions. Each gluon TMD contains
two gauge links in contrast to the quark TMDs that contain one. Because of
this, the process dependence of gluon TMDs is more involved than quark TMDs
\cite{Buffing:2013kca}. 
The simplest possible configurations are both future pointing [++] or one
future and one past pointing [+-] gauge links. In the literature related to 
small-$x$ physics, the former is called Weizs\"acker-Williams (WW) gluon
distribution \cite{Kovchegov:1998bi,McLerran:1998nk}.  For unpolarized gluons, 
WW gluon distribution can be
interpreted as the number density of gluons inside hadrons in light-cone
gauge. The other distribution, [+-], is called dipole distribution \cite{
Dominguez:2011br}. This
appears in many physical processes and is the Fourier transform of the color
dipole amplitude \cite{Iancu:2002xk, Kharzeev:2003wz}. In small 
$x$ physics, these two types of unintegrated gluon
distributions have been discussed in the literature quite extensively
\cite{Boer:2016xqr,Dominguez:2010xd,Metz:2011wb,Schafer:2013wca,
Akcakaya:2012si,Dumitru:2015gaa,Marquet:2017xwy,Dominguez:2011wm,Boer:2017xpy}. 
Apart from the unpolarized gluon TMD, the linearly polarized gluon TMD 
recently has attracted quite a lot of interest. This  basically measures an
interference between an amplitude when the active gluon is polarized along $x$
(or $y$) direction and a complex conjugate amplitude with the gluon polarized in
$y$ (or $x$) direction in an unpolarized hadron \cite{Dominguez:2011br}.
This was introduced for the first time in \cite{Mulders:2000sh}
 and calculated in a model in \cite{Meissner:2007rx}. It has been shown that the 
linearly polarized gluon distribution affect the unpolarized cross section
of scattering processes, as well an azimuthal asymmetry of the type $cos
2 \phi_h$ \cite{Pisano:2013cya} . The linearly polarized gluon distribution is a time-reversal even
(T even) object, and can be WW type or dipole type, depending on the gauge
links. $J/\psi$ production in $ep$ collision probes the WW type linearly
polarized gluon TMD through a virtual photon-gluon fusion process. The
leading order (LO) process $\gamma^*+g \rightarrow c {\bar c} \rightarrow
J/\psi$ contributes to the asymmetry at $z=1$ \cite{Mukherjee:2016qxa}. The linearly polarized
gluon distribution has not been extracted from data yet. However, there are
quite a large amount of theoretical studies about how to probe it in
different experiments. In \cite{Boer:2009nc} the authors have proposed to probe it in dijet
imbalance in the unpolarized hadronic collision  and also in heavy quark pair
production in ep collision and in pp collision
\cite{Pisano:2013cya,Marquet:2017xwy, Efremov:2017iwh, Efremov:2018myn}. It
can also be probed in quarkonium pair production in $pp$ collision
\cite{Lansberg:2017dzg}, and in associated production of a dilepton and
$J/\psi$ \cite{Lansberg:2017tlc}. Very
recently, in \cite{Dumitru:2018kuw} the authors have investigated the
possibility to probe it in dijet imbalance in $eA$ collision.  
$h_1^{\perp g}$ affects the transverse momentum distribution of final state
hadron like Higgs boson \cite{Sun:2011iw, Boer:2013fca,Boer:2011kf, Echevarria:2015uaa} and 
heavy quarkonium \cite{Boer:2012bt,Mukherjee:2016cjw, Mukherjee:2015smo}
 in unpolarized pp collision. Although $h_{1}^{\perp g} $ can be probed in $pp$ and $pA$
collision, initial and final state interactions may affect the factorization
in such processes. Such complications are less in $ep$ collision processes
for example at the electron-ion collider (EIC). In a previous work \cite{Mukherjee:2016qxa} we have
investigated the possibility of probing $h_{1}^{\perp g} $ in $cos~2 \phi$
asymmetry in $J/\psi $ production through the leading order (LO) process 
$ \gamma^*+g \rightarrow J/\psi$ at the future EIC. This $2 \rightarrow 1$
process contributes at $z=1$, where $z$ is the energy fraction of the photon
carried by the $J/\psi$ in the proton rest frame. Here, we extend our
analysis to the kinematical region $z<1$. We consider  the unpolarized $eP$ 
collision. The production mechanism of $J/\psi$ is not yet well-understood theoretically.
The most widely used approach is based on non-relativistic QCD (NRQCD)
\cite{Bodwin:1994jh}.
 Here one assumes a factorization of the amplitude into a hard part where
the $c {\bar c}$ pair is produced perturbatively in the process $\gamma^* +g 
\rightarrow c + {\bar c}+g$. The heavy quark pair then hadronizes to form the
$J/\psi$ bound state. The hadronization is described in terms of the long
distance matrix elements (LDMEs) which are obtained by fitting the data. For
some LDMEs lattice calculations are available.  They have definite scaling
properties with respect to the velocity parameter $v$, which is assumed to
be small. The cross section for the production of $J/\psi$ is expressed as a
double expansion  in terms of the strong coupling constant $\alpha_s$ as
well as $v$ \cite{Lepage:1992tx}. For $J/\psi$, $v \approx 0.3$. In NRQCD, the heavy quark pair
can be produced both in color singlet (CS) state \cite{Carlson:1976cd, Berger:1980ni, Baier:1981uk,
Baier:1981zz} or in color octet (CO) state \cite{Braaten:1994vv,Cho:1995vh, Cho:1995ce}.
The former is called CS model and the latter, CO model. In the CS
model, the heavy quark pair is produced in the hard process as a color 
singlet with the same quantum number as $J/\psi$. In \cite{DAlesio:2017rzj}
the $J/\psi$ production rate for unpolarized $pp$ collision at RHIC assuming a
generalized TMD factorization was calculated in CS model, and it was found
that the theoretical estimate reasonably explains the data for low values of
$p_T$, where $p_T$ is the transverse momentum of $J/\psi$. However, high
$p_T$ spectra for $J/\psi$ production needs the inclusion of CO states. 
As we showed in \cite{Rajesh:2018qks} both CS and CO contributions are needed to match the HERA
data. However, in this work, as a first study, we calculate the $cos ~
2 \phi$  asymmetry in $J/\psi$ production in $ep$ collision in CS model.  All previous studies of this 
asymmetry in $eP$ collision have  considered the LO process. In this work, for the first time, we
investigate the asymmetry in the kinematical region $z <1$.  As we are interested in small x region, 
we consider the process $ \gamma^*+g \rightarrow J/\psi+g$, as gluon distributions are dominant at 
small $x$. This process probes the  WW type gluon TMDs.

In order to estimate the $cos ~ 2 \phi$ asymmetry, we use three different models for the TMDs. 
First, we use a Gaussian parametrization \cite{Boer:2012bt,
Mukherjee:2015smo, Mukherjee:2016cjw} for both the linearly polarized gluon distribution and the unpolarized TMD.
The linearly polarized gluons satisfy an upper bound and the asymmetry reaches its maximum value when this upper bound is 
saturated. We also calculate the upper bound of the asymmetry. Finally, in the small $x$ region, the WW type gluon distributions 
are calculated  using a saturation model \cite{McLerran:1993ni,
McLerran:1993ka, McLerran:1994vd}.  TMDs in McLerran-Venugopalan (MV) model, although  expected to work better
for a large nucleus, has been found to be phenomenologically successful for the nucleon
\cite{Bacchetta:2018ivt}. We have used a regulated MV model in small $x$ region for the WW type gluon TMDs. We have compared
 the asymmetry in all three cases in the kinematics of the planned electron-ion collider (EIC). 
 
 The paper is organized into six sections starting with the introduction in Sec.~I. In Sec.~II, we provide the analytic framework, kinematics of the process and the calculations of asymmetry in different models. We provide the numerical estimations in Sec.~III and conclude the results in Sec.~IV. Some detailed analytic results are given in the appendix. 

\section{Framework for calculation}

The process we have considered here is
\be
e(l)+p(P)\rightarrow e(l')+J/\psi (P_h)+X
\ee

 Both the scattering electron and target proton are unpolarized. Four momentum of particles is represented within the round 
brackets. The dominating subprocess for small $x$ for quarkonium production in $ep$ collision is photon-gluon fusion process, at leading 
order this process contributes at $z=1$ \cite{Mukherjee:2016qxa}.  In this work, we consider the NLO process 
  $\gamma^*(q)+g(k)\rightarrow J/\psi(P_h) +g(p_g)$ and the kinematical region $z<1$, which will be accessible at EIC. 
The final state gluon is not detected. Here the variable $z$ is defined as $z=P\cdot P_h/P\cdot q$ which is the energy 
fraction of $J/\psi$ in the proton rest frame.  We use a generalized factorization scheme taking into account the partonic 
transverse momenta. We consider the frame in which the virtual photon and proton are moving in $+z$ and $-z$ direction respectively.
 The incoming and outgoing electron form a lepton plane, which provides a reference for measuring azimuthal angles of other 
particles. The four momenta of proton and virtual photon $q=l-l'$ are  given by
\cite{Mukherjee:2016qxa}:
\be
P=n_- + \frac{M_p^2}{2}n_+ \approx n_- \\
q=-x_Bn_-+\frac{Q^2}{2x_B}n_+\approx -x_BP+(P\cdot q)n_+
\ee 
where $Q^2=-q^2$ and Bjorken variable, $x_B=\frac{Q^2}{2P\cdot q}$. $M_p$ is the mass of proton. 
All four momenta are written in terms of light like vectors $n_-=P$ and $n_+=n=(q+x_BP)/P\cdot q$, such that $n_+\cdot n_-=1$ and $n_-^2=n_+^2=0$. The leptonic momenta can be written as
\be
l=\frac{1-y}{y}x_BP+\frac{1}{y}\frac{Q^2}{2x_B}n+\frac{\sqrt(1-y)}{y}Q\hat{l}_{\perp}=\frac{1-y}{y}x_BP+\frac{s}{2}n+\frac{\sqrt(1-y)}{y}Q\hat{l}_{\perp}\\
l'=\frac{1}{y}x_BP+\frac{1-y}{y}\frac{Q^2}{2x_B}n+\frac{\sqrt(1-y)}{y}Q\hat{l}_{\perp}=\frac{1}{y}x_BP+(1-y)\frac{s}{2}n+\frac{\sqrt(1-y)}{y}Q\hat{l}_{\perp}
\ee 
here, $s=(l+P)^2=2P\cdot l$, is the center of mass energy of electron-proton scattering. $y=P\cdot q/P\cdot l$, 
such that the relation $Q^2=sx_By$ hold. The virtual photon and target proton system invariant mass squared is defined as $W^2=(P+q)^2$. In terms of the light-like vectors defined above, the four momenta of initial state gluon is given as
\be
k=xP+k_{\perp}+(p\cdot P-xM_p^2)n\approx xP+k_{\perp}
\ee
where, $x=k\cdot n$ is the light-cone momentum fraction. The four momentum of the final state $J/\psi$ and
the final state gluon are give by
\be
P_h=z(P\cdot q)n+\frac{M^2+\textbf{P}_{h\perp}^2}{2zP\cdot q}P+P_{h\perp}\\
p_g=(1-z)(P\cdot q)n+\frac{\textbf{p}_{g\perp}^2}{2(1-z)P\cdot q}P+p_{g\perp}
\ee
$P_h^2=-\textbf{P}_{h\perp}^2$. $M$ is the mass of $J/\psi$. \par

For the partonic level process: $\gamma^*(q)+g(k)\rightarrow J/\psi(P_h)+g(p_g)$, we can define the Mandelstam variables as follows
\be
\hat{s}=(k+q)^2=q^2+2k\cdot q=\frac{xQ^2}{x_B}-Q^2,
\label{xb}
\ee
\be
\hat{t}&=&(k-P_h)^2=M^2-2k\cdot P_h\nonumber\\
&=&M^2-\frac{xzQ^2}{x_B}+2k_{\perp }P_{h\perp}\cos(\phi-\phi_h),
\ee
\be
\hat{u}&=&(q-P_h)^2=M^2+q^2-2q\cdot P_h\nonumber\\
&=&M^2-(1-z)Q^2-\frac{M^2+P^2_{h\perp}}{z}.
\ee
The $\phi$ and $\phi_h$ are the azimuthal angles of the initial gluon and $J/\psi$ transverse momentum vector respectively.\\

We use a framework based on generalized parton model approach with the inclusion of intrinsic transverse momentum effects, and assume TMD factorization.  The differential cross section for the unpolarized process 
is given by \cite{Mukherjee:2016qxa} ;
\be
\begin{aligned}
d\sigma=&\frac{1}{2s}\frac{d^3l'}{(2\pi)^32E_{l'}}\frac{d^3P_h}{(2\pi)^32E_{P_h}}\int \frac{d^3p_g}{(2\pi)^32E_g}\int dx 
d^2{\bm k}_{\perp}(2\pi)^4\delta(q+k-P_h-p_g)\\
&\times\frac{1}{Q^4}L^{\mu\mu'}(l,q)\Phi^{\nu\nu'}(x,{\bm k}_{\perp})\mathcal{M}^{\gamma^*+g\rightarrow J/\psi +g}_{\mu\nu}\mathcal{M}^{*\gamma^*+g\rightarrow J/\psi +g}_{\mu'\nu'}
\end{aligned}
\ee

where $L^{\mu\mu'}$ is leptonic tensor which is given by
\be
L^{\mu\mu'}(l,q)=e^2(-g^{\mu\mu'}Q^2+2(l^{\mu}l'^{\mu'}+l^{\mu'}l'^{\mu}))
\ee
with $e$ is the electric charge of electron. \\
$\Phi^{\nu\nu'}$ is gluon correlator which can be parametrized in terms of gluon TMDs. For unpolarized proton, 
at leading twist, gluon correlator can be given as \cite{Mulders:2000sh}: 
\be
\phi^{\nu\nu'}_g(x,\textbf{k}_{\perp})=\frac{1}{2x}[-g^{\nu\nu'}_{\perp}f_1^g(x,\textbf{k}_{\perp}^2)+(\frac{k_{\perp}^{\nu}k_{\perp}^{\nu'}}{M_p^2}+g_{\perp}^{\nu\nu'}\frac{\textbf{k}_{\perp}^2}{2M_p^2})h_1^{\perp g}(x,\textbf{k}_{\perp}^2)]
\ee
  
where $f_1^g(x,{\bm k}_{\perp}^2)$ is the unpolarized gluon distribution and 
$h_1^{\perp g}(x,{\bm k}_{\perp}^2)$ is the linearly polarized gluon distribution. 
$g_{\perp}^{\nu\nu'}=g^{\nu\nu'}-P^{\nu}n^{\nu'}/P\cdot n-P^{\nu'}n^{\nu}/P\cdot n$.

\section*{$J/\psi$ production in NRQCD based color singlet (CS) framework}

\begin{figure}[H]
	\begin{minipage}[c]{0.99\textwidth}
	    \hspace{1.5cm}
		\includegraphics[width=14cm,height=6.5cm,clip]{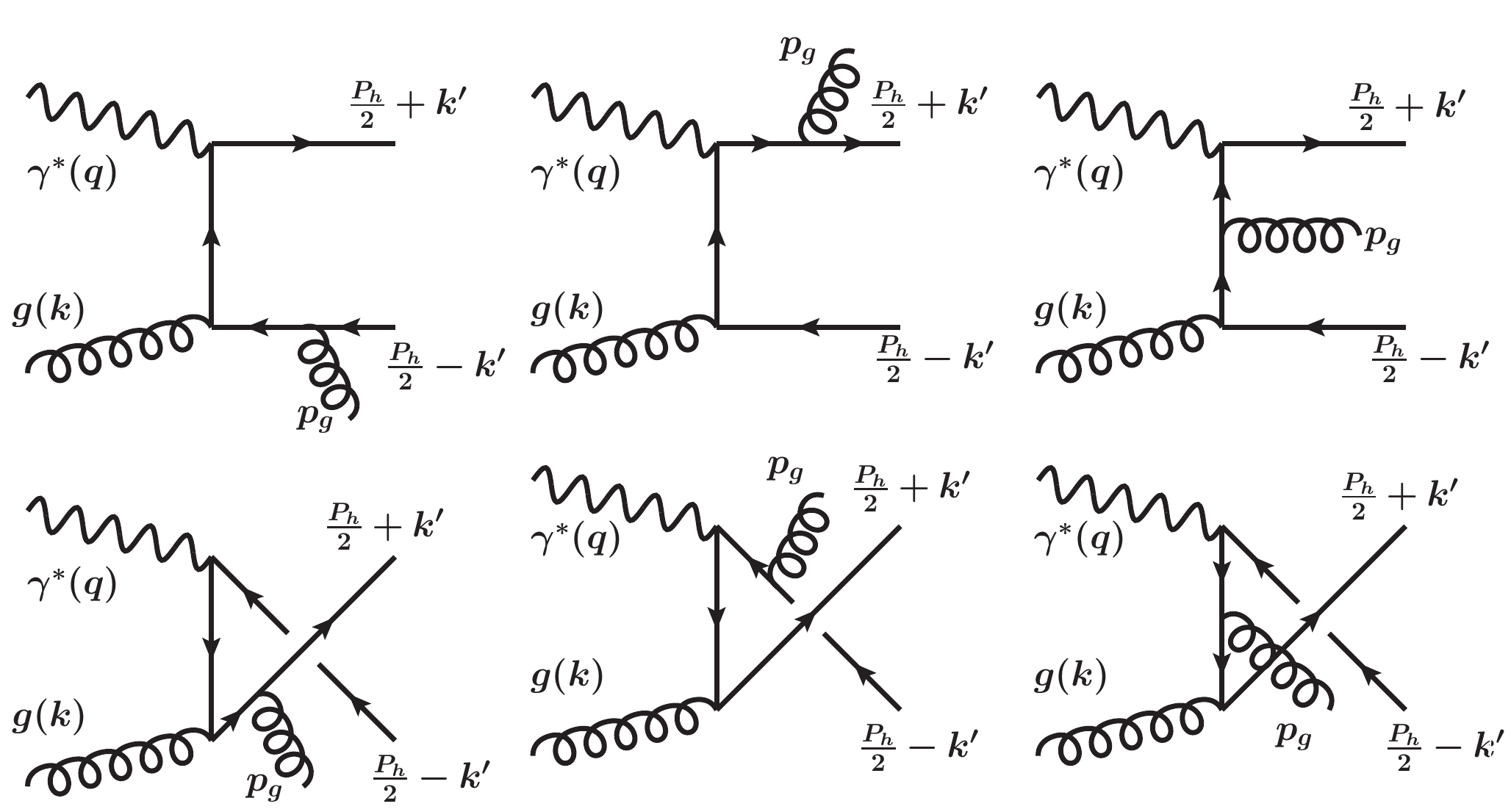}
	\end{minipage}
	\caption{\label{fig3}  Feynman diagrams for $\gamma^*+g\rightarrow J/\psi +g$ process}
\label{figure1}
\end{figure}

The dominating subprocess at  is $\gamma^*+g\rightarrow J/\psi +g$. All the tree level Feynman diagrams corresponding to this process are given in Fig. \ref{figure1} . 

The general expression of the amplitude for the bound state production of $J/\psi$ in NRQCD framework can be written as 
\cite{Boer:2012bt, Mukherjee:2016qxa} : 

\begin{equation}
\begin{aligned}
\mathcal{M}\left(\gamma^* g\rightarrow Q\bar{Q}[\leftidx{^{2S+1}}{L}{_J}^{(1)}](P_h)+g\right)=\sum_{L_zS_z}\int\frac{d^3\bm{k}^{\prime}}{(2\pi)^3}\Psi_{LL_z}(\bm{k}^\prime) \langle LL_z;SS_z|JJ_z\rangle \\
\times\mathrm{Tr}[O(q,k,P_h,k^\prime)\mathcal{P}_{SS_z}(P_h,k')],
\end{aligned}
\end{equation}

As we have imposed a cutoff on $z$, $ z < 0.9$, we do not need to consider the virtual diagrams as they contribute at $z=1$. 
In the above equation, $2k'$ is the relative momentum of heavy quarks and $O(q,k,P_h,k^\prime)$ is calculated from the Feynman diagrams. The spinors of heavy quark, anti-quark legs are absorbed into the bound state  wave function. By considering contribution from all the Feynman diagrams, $O(q,k,P_h,k^\prime)$ is given by
\begin{equation}
O(q,k,P_h,k^\prime)=\sum_{m=1}^6 \mathcal{C}_m O_m(q,k,P_h,k^\prime).
\end{equation}
Where, $O_m,\ (m=1,2,...6)$ are corresponding to each Feynman diagrams and $\mathcal{C}_m$ represents
the color factor of corresponding diagram. The expressions for $O_m$ are given below

\begin{equation}\label{ee1}
\begin{aligned}
O_1= 
4g^2_s(ee_c)\varepsilon^{\rho\ast}_{\lambda_g}
(p_g)
\gamma_\nu\frac{\slashed{P_h}+2\slashed{k}^\prime-2\slashed{q}+M}{(P_h+2k^\prime-2q)^2-M^2}\gamma_\mu
\frac{-\slashed{P_h}+2\slashed{k}^\prime-2\slashed{p}_g+M}{(P_h-2k^\prime+2p_g)^2-M^2}\gamma_\rho,
\end{aligned}
\end{equation}
\begin{equation}\label{ee2}
\begin{aligned}
O_2= 
4g^2_s(ee_c)\varepsilon^{\rho\ast}_{\lambda_g}
(p_g)
\gamma_\rho\frac{\slashed{P_h}+2\slashed{k}^\prime+2\slashed{p}_g+M}{(P_h+2k^\prime+2p_g)^2-M^2}\gamma_\nu
\frac{-\slashed{P_h}+2\slashed{k}^\prime+2\slashed{k}+M}{(P_h-2k^\prime-2k)^2-M^2}\gamma_\mu,
\end{aligned}
\end{equation}
\begin{equation}\label{ee3}
\begin{aligned}
O_3= 
4g^2_s(ee_c)\varepsilon^{\rho\ast}_{\lambda_g}
(p_g)
\gamma_\nu\frac{\slashed{P_h}+2\slashed{k}^\prime-2\slashed{q}+M}{(P_h+2k^\prime-2q)^2-M^2}\gamma_\rho
\frac{-\slashed{P_h}+2\slashed{k}^\prime+2\slashed{k}+M}{(P_h-2k^\prime-2k)^2-M^2}\gamma_\mu,
\end{aligned}
\end{equation}

Here, the mass of bound state $M$ is assumed to be twice the mass of charm quark ($m_c$)  $i.e.\  M=2m_c$, .The charge conjugation invariance allow us to write the expressions for ($O_4,O_5$ and $O_6$), from the other Feynman diagrams,  by reversing the fermion line and replacing $k'$ by $-k'$. Assuming the $Q\bar{Q}$ is formed in color singlet state, the color factor of each diagram is given by
\begin{equation}
\begin{aligned}
\mathcal{C}_1=\mathcal{C}_5=\mathcal{C}_6= \sum_{ij}\langle 
3i;\bar{3}j|1\rangle(t_at_b)_{ij},~~~\mathcal{C}_2=\mathcal{C}_3=\mathcal{C}_4= \sum_{ij}\langle
3i;\bar{3}j|1\rangle(t_bt_a)_{ij}
\end{aligned}
\end{equation} 	
The SU(3) Clebsch-Gordan coefficients for CS are given by
\begin{eqnarray}
\langle 3i;\bar{3}j|1\rangle=\frac{\delta^{ij}}{\sqrt{N_c}}
\end{eqnarray}
where $N_c$ is the number of colors. The generators of the SU(3) group satisfies the relations: 
$\mathrm{Tr}(t_a)=0,$ $\mathrm{Tr}(t_at_b)=\delta_{ab}/2$ and 
$\mathrm{Tr}(t_at_bt_c)=
\frac14(d_{abc}+if_{abc})$.

From these relations, we get the color factor for the production of $Q\bar{Q}$ pair in CS state as follows;
\be
\mathcal{C}_1=\mathcal{C}_2=\mathcal{C}_3=\mathcal{C}_4=\mathcal{C}_5=\mathcal{C}_6=\frac{\delta_{ab}}{2\sqrt{N_c}}.
\ee
The spin projection operator, given in the equation of amplitude of the bound state, includes the spinors of heavy quark 
and anti-quark and is given by \cite{Boer:2012bt}: 
\begin{eqnarray}
\mathcal{P}_{SS_z}(P_h,k^\prime)&=&\sum_{s_1s_2}\langle\frac12s_1;\frac12s_2|SS_z\rangle v(\frac{P_h}{2}-
k^\prime,s_1)\bar{u}(\frac{P_h}{2}+k^\prime,s_2)\nonumber\\
&=&\frac{1}{4M^{3/2}}(-\slashed{P}_h+2\slashed{k}^\prime+M)\Pi_{SS_z}(\slashed{P}_h+2\slashed{k}^\prime+M)
+\mathcal{O}(k^{\prime 2}) 
\end{eqnarray}
where  $\Pi_{SS_z}=\gamma^{5}$ for singlet ($S=0$) state and $\Pi_{SS_z}=\slashed{\varepsilon}_{s_z}(P_h)$
for triplet ($S=1$) state. $\varepsilon_{s_z}(P_h)$ is the spin polarization vector of $Q\bar{Q}$ pair. Since, $k'<<P_h$, one can  perform Taylor expansion of the amplitude around $k'=0$. In that expansion, the first term gives the S-waves(L=0,J=0,1). For the P-waves(l=1,J=0,1,2), we need to consider the linear terms in $k'$ in the expansion as the radial wavefunction $R_1(0)=0$ for $P-$wave.  Since, $J/\psi$ is a $^3S_1$ state, in the color singlet model we calculate contribution of the CS state $^3S_1$.

\begin{eqnarray}\label{e8}
\mathcal{M}[\leftidx{^{2S+1}}{S}{_J}^{(1)}](P_h,k)&=&\frac{1}{\sqrt{4\pi}}R_0(0)\mathrm{Tr}[O(q,k,P_h,
k^\prime) \mathcal{P}_{SS_z}(P_h,k^\prime)]\Big\rvert_{k^\prime=0}\nonumber\\
&=&\frac{1}{\sqrt{4\pi}}R_0(0)\mathrm{Tr}[O(0) \mathcal{P}_{SS_z}(0)],
\end{eqnarray}
where,
\be
O(0)=O(q,k,P_h,k^\prime)\Big\rvert_{k^\prime=0}~,~~~~~~~~  
\mathcal{P}_{SS_z}(0)=\mathcal{P}_{SS_z}(P_h,k^\prime)\Big\rvert_{k^\prime=0}
\ee

We have the following symmetry relations for ${^3}{S}{_1}$ state
\begin{eqnarray}\label{e15}
\mathrm{Tr}[O_1(0)(-\slashed{P}_h+M)\slashed{\varepsilon}_{s_z}]&=&
\mathrm{Tr}[O_4(0)(-\slashed{P}_h+M)\slashed{\varepsilon}_{s_z}]\nonumber\\
\mathrm{Tr}[O_2(0)(-\slashed{P}_h+M)\slashed{\varepsilon}_{s_z}]&=&
\mathrm{Tr}[O_5(0)(-\slashed{P}_h+M)\slashed{\varepsilon}_{s_z}]\nonumber\\
\mathrm{Tr}[O_3(0)(-\slashed{P}_h+M)\slashed{\varepsilon}_{s_z}]&=&
\mathrm{Tr}[O_6(0)(-\slashed{P}_h+M)\slashed{\varepsilon}_{s_z}]\nonumber\\
\end{eqnarray}
The final expression for the amplitude for ${^3}{S}{_1}$ state  is given by
\begin{eqnarray}\label{e16}
\begin{aligned}
\mathcal{M}[\leftidx{^{3}}{S}{_1}^{(1)}](P_h,k)=&{}\frac{1}{4\sqrt{\pi M}}R_0(0)\frac{\delta_{ab}}{\sqrt{N_c}}
\mathrm{Tr}\left[\sum_{m=1}^3O_m(0)(-\slashed{P}_h+M)\slashed{\varepsilon}_{s_z}\right],
\end{aligned}
\end{eqnarray}
where
\begin{eqnarray}\label{e17}
\begin{aligned}
\sum_{m=1}^3O_m(0)=&{}g^2_s(ee_c)\varepsilon^{
	\rho\ast}_{
	\lambda_g}(p_g)\Bigg[\frac{\gamma_\nu(\slashed{P_h}-2\slashed{q}+M)\gamma_\mu
	(-\slashed{P_h}-2\slashed{p}_g+M)\gamma_\rho}{(\hat{s}-M^2)(\hat{u}-M^2+q^2)}\\&
\qquad\qquad+\frac{\gamma_\rho(\slashed{P_h}+2\slashed{p}_g+M)\gamma_\nu
	(-\slashed{P_h}+2\slashed{k}+M)\gamma_\mu}{(\hat{s}-M^2)(\hat{t}-M^2)}\nonumber\\&\qquad\qquad+
\frac{\gamma_\nu(\slashed{P_h}-2\slashed{q}+M)\gamma_\rho(-\slashed{P_h}+2\slashed{k}+M)\gamma_\mu
}{(\hat{t}-M^2)(\hat{u}-M^2+q^2)}
\Bigg].
\end{aligned}
\end{eqnarray}  

\section*{Calculation of the asymmetry}

We use a framework based on generalized parton model, with the
inclusion of intrinsic transverse momentum effects. We assume TMD factorization for
the process considered. We consider a kinematical region in which the transverse momentum 
of $J/\psi$ is small compared 
to the mass of $J/\psi$, $M$ $i.e. ~~ P_{h\perp}<M$.
The final gluon carries the momenta fraction $(1-z)$, as $z=\frac{P\cdot P_h}{P\cdot q}$, is the energy fraction transferred from the photon to $J/\psi$ in the proton rest frame. So, this means that when $~~z\rightarrow 1$, the outgoing gluon is soft. We consider $z<0.9$ to keep the final gluon hard. Moreover, gluon and heavy quark  fragmentation can also contribute to quarkonium production significantly for $P_{h\perp}>4~GeV$. We have imposed an upper limit on $P_{h\perp}$. In order to eliminate the fragmentation of the hard gluon into $J/\psi$ we also use a lower bound on $z$, namely  $0.3<z$. \\

In the differential cross section given in Eq.~(12), there is a contraction of four tensors which is written as
\be
L^{\mu\mu'}(l,q)\Phi^{\nu\nu'}(x,{\bm k}_{\perp})\mathcal{M}^{\gamma^*+g\rightarrow J/\psi +g}_{\mu\nu}\mathcal{M}^{*\gamma^*+g\rightarrow J/\psi +g}_{\mu'\nu'}
\ee
where the individual components are defined above. The summation over the transverse polarization of the final on-shell gluon is given by
\be
\sum_{\lambda_a=1}^2\varepsilon^{\lambda_a}_\mu(p_g)\varepsilon^{\ast\lambda_a}_{\mu^\prime}(p_g)=-g_{
	\mu\mu^\prime}+
\frac{p_{g\mu} n_{g\mu^\prime}+p_{g\mu^\prime}n_{g\mu}}{p_g\cdot n_g}-\frac{p_{g\mu} p_{g\mu^\prime}}{(p_g\cdot n_g)^2}
\ee
with $n_g^\mu=\frac{P^\mu_h}{M}$.
We have three amplitudes and their corresponding conjugates, given by Eq.~(27), that will contribute. We use the notation 
\be
\begin{aligned}
	\mathcal{M}_i[\leftidx{^{3}}{S}{_1}^{(1)}](P_h,k)=&{}\frac{1}{4\sqrt{\pi M}}R_0(0)\frac{\delta_{ab}}{\sqrt{N_c}}
	\mathrm{Tr}\left[O_i(0)(-\slashed{P}_h+M)\slashed{\varepsilon}_{s_z}\right],
\end{aligned}
\ee
where $i=1,2,3$, corresponds to the contribution from each independent diagram. \\
So, the cross section will get contribution from  nine terms $( \mathrm{of~ the~ form} ~ M_iM_j~ ,~ \mathrm{where} ~ i,j=1,2,3)$
\be
M_iM_j=L^{\mu\mu'}(l,q)\Phi^{\nu\nu'}(x,{\bm k}_{\perp})\mathcal{M}^{\gamma^*+g\rightarrow J/\psi +g}_{i\mu\nu}\mathcal{M}^{*\gamma^*+g\rightarrow J/\psi +g}_{j\mu'\nu'}
\ee
and hence, the differential cross section can be written as 
\be
\begin{aligned}
	d\sigma=&\frac{1}{2s}\frac{d^3l'}{(2\pi)^32E_{l'}}\frac{d^3P_h}{(2\pi)^32E_{P_h}} \int \frac{d^3p_g}{(2\pi)^32E_g}\int dx 
	d^2{\bm k}_{\perp}(2\pi)^4\delta(q+k-P_h-p_g)\\
	&\times\frac{1}{Q^4} {\mid M \mid }^2;
\end{aligned}
\ee
where $M = \sum_i M_i$. Out of the nine terms in ${\mid M \mid}^2$, six are interference terms with a symmetry $M_iM_j=M_jM_i ~~for~~ i\ne j$.
So, effectively we need to compute six terms.\\

In a frame where the virtual photon and target proton move along the $z$-axis, and the lepton scattering plane defines the azimuthal angles $\phi_l=\phi_{l'}=0$, then we have 
\be
\frac{d^3l'}{(2\pi)^32E_{l'}}=\frac{1}{16\pi^2}sydx_Bdy, ~~~~ \frac{d^3P_h}{(2\pi)^32E_{h}}=\frac{1}{(2\pi)^3}\frac{1}{2z}dzd^2\textbf{P}_{h\perp}\nonumber\\
 \frac{d^3p_g}{(2\pi)^32E_{g}}=\frac{1}{(2\pi)^3}\frac{1}{2z_2}dz_2d^2\textbf{p}_{g\perp} 
\ee
and the delta function can be expressed as
\be
\begin{split}
\delta^4(q+k-P_h-p_g)=&\delta\Big ( x-\frac{1}{ys}(x_Bys+\frac{M^2+P_{h\perp}^2}{z}+\frac{(k_{\perp}-P_{h\perp})^2}{(1-z)}) \Big )\\
&\times\frac{2}{ys}\delta(1-z-z_2) \times\delta^2(\textbf{k}_{\perp}-\textbf{P}_{h \perp}-\textbf{p}_{g\perp})
\end{split}
\ee
where, the delta function sets $z_2=(1-z)$. Hence, after integrating with respect to  $x$, $z_2$ and $p_{g\perp}$, 
the final form of the differential cross section can be given by
\be
\frac{d\sigma}{dydx_Bdzd^2\textbf{P}_{h\perp}}=\frac{1}{256\pi^4}\frac{1}{x_B^2s^3y^2z(1-z)}\int k_{\perp}dk_{\perp}
{\mid M' \mid }^2
\ee

where, ${\mid M' \mid }^2=\int d\phi{\mid M \mid }^2$, and $k_\perp$ is the
magnitude of ${\bf k_\perp}$.
As we are interested in the small $x$ region, we neglect terms containing higher powers of  $x_B$; also as $z<1$, we neglect  terms containing higher powers of $z$ and kept up to $z^2$. We also keep terms only up to  $({k^2_\perp\over M^2_p})$. 
The leading terms in the numerator of the $cos(2\phi_h)$ asymmetry come only from the first Feynman diagram.
These terms are given in the appendix.
The denominator of the $cos(2\phi_h)$ asymmetry, which is defined below, is simply the cross section integrated over the azimuthal angle $\phi_h$. The leading terms in the cross section comes from $f_1^g$ term. All the terms corresponding to $h_1^{\perp g}$ are suppressed by $k_{\perp}^2/M_p^2$. Hence, from the approximations we mentioned above, the leading terms in the cross section in the denominator of $cos(2\phi_h)$ asymmetry are coming from $M_1M_1,~M_1M_2,~M_1M_3$. These contributions are given in the appendix.\\


The differential cross section then can be given as
\begin{align} \label{csf}
\frac{d\sigma}{dydx_Bdzd^2\textbf{P}_{hT}}&=\frac{1}{256\pi^4}\frac{1}{x_B^2s^3y^2z(1-z)}\int k_{\perp}dk_{\perp}\nonumber\\&
\{(A_0+A_1cos\phi_h)f_1^g(x,\textbf{k}_{\perp}^2) \}+
\frac{k_{\perp}^2}{M_p^2}\{(B_0+B_1cos\phi_h+B_2cos2\phi_h)h_1^{\perp g}(x,\textbf{k}_{\perp}^2) \}
\end{align}

The coefficients $A_0, A_1, B_0, B_1$ and $B_2$ are given in the appendix. The $cos(2\phi)$ asymmetry is defined as 
\be
\langle cos(2\phi_h)\rangle=\frac{\int d\phi_hcos(2\phi_h)d\sigma}{\int d\phi_h d\sigma}
\ee
In order to estimate the $cos(2\phi_h)$ asymmetry, we need to parametrize the TMDs. We discuss two parametrization models, Gaussian parameterization 
of the TMDs and McLerran-Venugopalan(MV) model. We also calculate the upper bound of the asymmetry.

\subsection {Gaussian parametrization of  the TMDs}

Both for the linearly polarized gluon distribution and the unpolarized gluon TMD, a Gaussian parametrization is 
used widely in the literature. The linearly polarized gluon distribution satisfies  the model independent positivity bound 
\cite{Mulders:2000sh};

\be
\frac{\textbf{k}_{\perp}^2}{2M_p^2}\left|h_1^{\perp g}(x,\textbf{k}_{\perp}^2)\right| \leq f_1^g(x,\textbf{k}_{\perp}^2)
\label{positivity}
\ee
The Gaussian parametrizations satisfy the positivity bound but does not saturate it. They are as follows
\cite{Boer:2012bt,Mukherjee:2015smo, Mukherjee:2016cjw}; 
\be
f_1^g(x,\textbf{k}_{\perp}^2)=f_1^g(x,\mu)\frac{1}{\pi\langle k_{\perp}^2\rangle}e^{-k_{\perp}^2/\langle k_{\perp}^2\rangle}
\ee
\be
h_1^{\perp g}(x,\textbf{k}_{\perp}^2)=\frac{M_p^2f_1^g(x,\mu)}{\pi\langle k_{\perp}^2\rangle^2}\frac{2(1-r)}{r}e^{1-\frac{k_{\perp}^2}{r\langle k_{\perp}^2\rangle}}
\ee         
where, $r(0<r<1)$ is a parameter, in our numerical estimates we took $r=1/3$.  $f_1^g(x, \mu)$ is the gluon collinear PDF, which is measured at the scale $\mu=\sqrt{M^2+P_{h\perp}^2}$ and it obeys the Dokshitzer-Gribov-Lipatov-Altarelli-Parisi (DGLAP) scale evolution. The width of the Gaussian, $\langle k_\perp^2 \rangle$, depends on the energy scale of the process.
Following \cite{ Boer:2012bt},  we took  $\langle k_\perp^2 \rangle=0.25 ~ \mathrm{GeV}^2$.  
The asymmetry increases on increasing model parameter $r$, reaches a maximum at $r\approx 0.4$ and then decreases, 
but the variation of asymmetry is very small.

\subsection {Upper bound of the asymmetry}
 
 The asymmetry reaches its maximum value when the positivity bound given by Eq. (\ref{positivity}) is saturated. 
 Using this, we calculate the upper bound  of $|\langle cos(2\phi_h)\rangle|$ as below
\cite{Pisano:2013cya}; 
\begin{align}
|\langle cos(2\phi_h)\rangle|&=\left| \frac{\int d\phi_hcos(2\phi_h)d\sigma}{\int d\phi_h d\sigma}\right|\nonumber\\&
=\frac{\int k_{\perp}dk_{\perp}\textbf{k}_{\perp}^2|h_1^{\perp g}(x,\textbf{k}_{\perp}^2)|}{\int k_{\perp}dk_{\perp}2M_pf_1^g(x,\textbf{k}_{\perp}^2)}\frac{|B_2|}
{A_0}\leq\frac{|B_2|}{A_0}\equiv R
\end{align}

\begin{figure}[H]
	\begin{minipage}[c]{0.99\textwidth}
		\small{(a)}\includegraphics[width=7cm,height=6.5cm,clip]{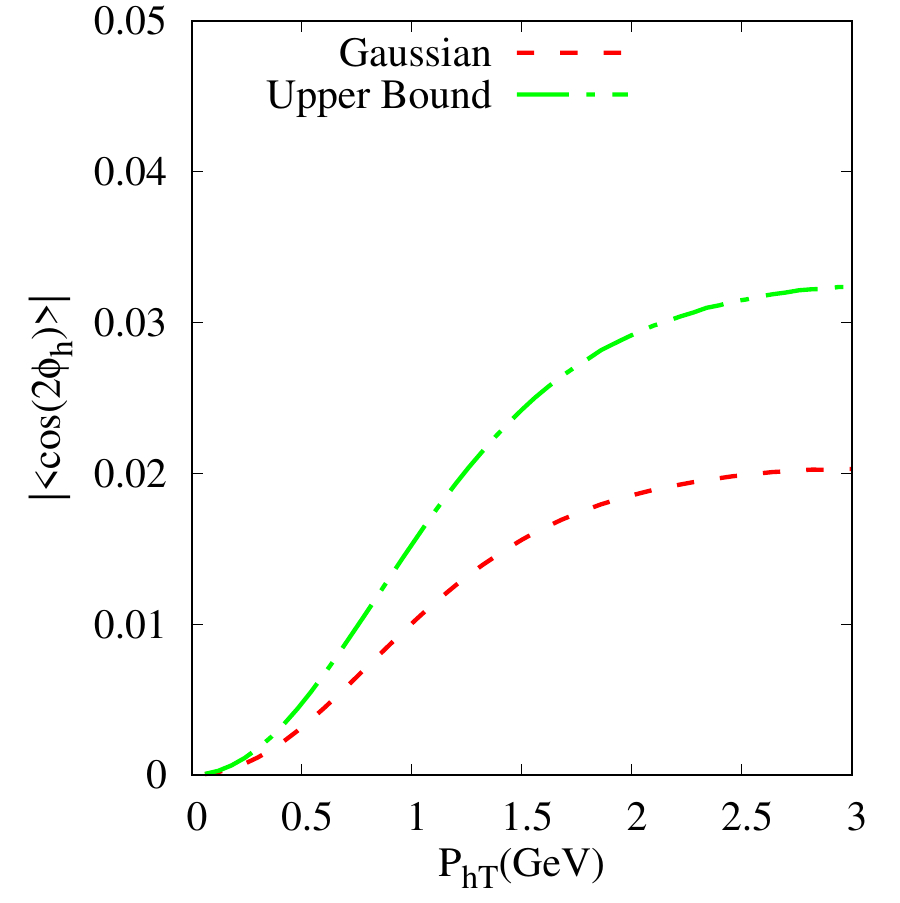}
		\hspace{0.1cm}
		\small{(b)}\includegraphics[width=7cm,height=6.5cm,clip]{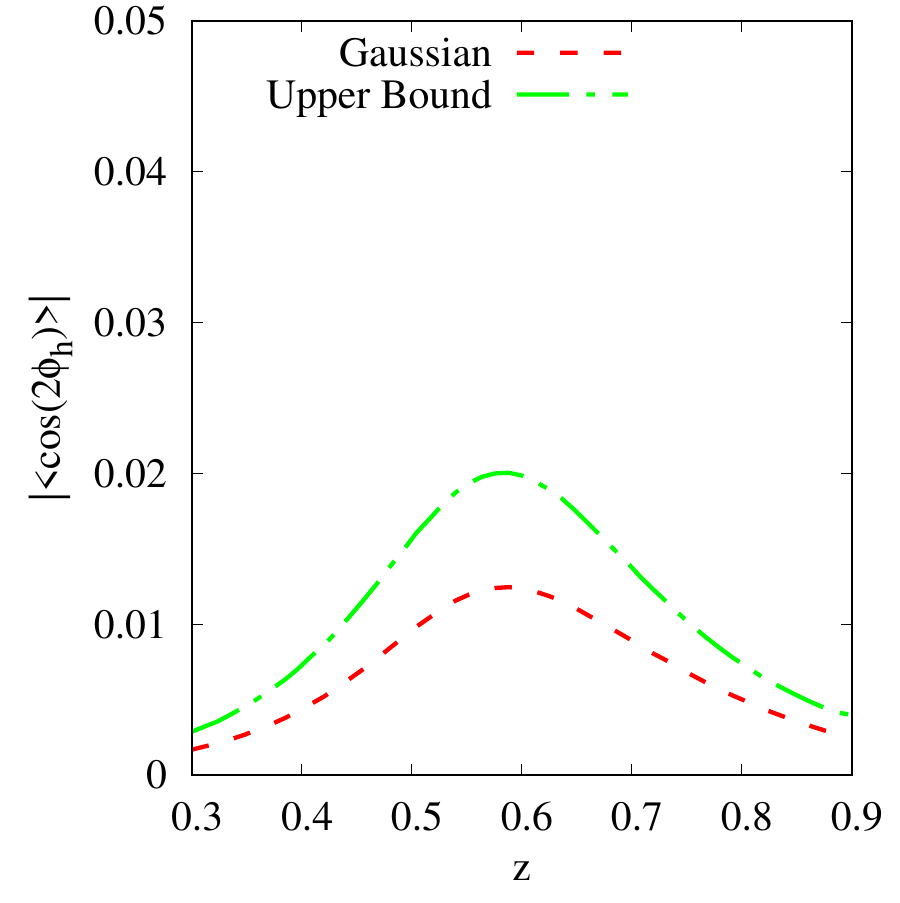}
	\end{minipage}
	\caption{\label{fig3}  $cos(2\phi_h)$ asymmetry  in $e+p\rightarrow e+ J/\psi +X$
		process as function of 
		(a) $P_{hT}$ (left panel) and  (b) $z$ (right panel) at $\sqrt{s}=45$ GeV (EIC) and $x_B=0.01$. The integration ranges are 
		$0<P_{hT}\leq3$ GeV, $0.3<z<0.9$ and $0.05<y<0.4$. For convention of lines see the legend in the plots.}
\label{figure2}
\end{figure}

\subsection{McLerran-Venugopalan (MV) model}

In the  small $x$ region, the  Weizs\"{a}cker-Williams (WW)  type gluon distribution can be calculated  in MV model
\cite{McLerran:1993ni, McLerran:1993ka, McLerran:1994vd}: 
Within the nonperturbative McLerran-Venugopalan model, we can define the gluon distribution function inside an unpolarized large nucleus or inside an energetic proton, in the small $x$ limit. In this model, the analytical expression of the WW type unpolarized and linearly polarized gluon distributions are given by 

\begin{figure}[H]
	\begin{minipage}[c]{0.99\textwidth}
		\small{(a)}\includegraphics[width=7cm,height=6.5cm,clip]{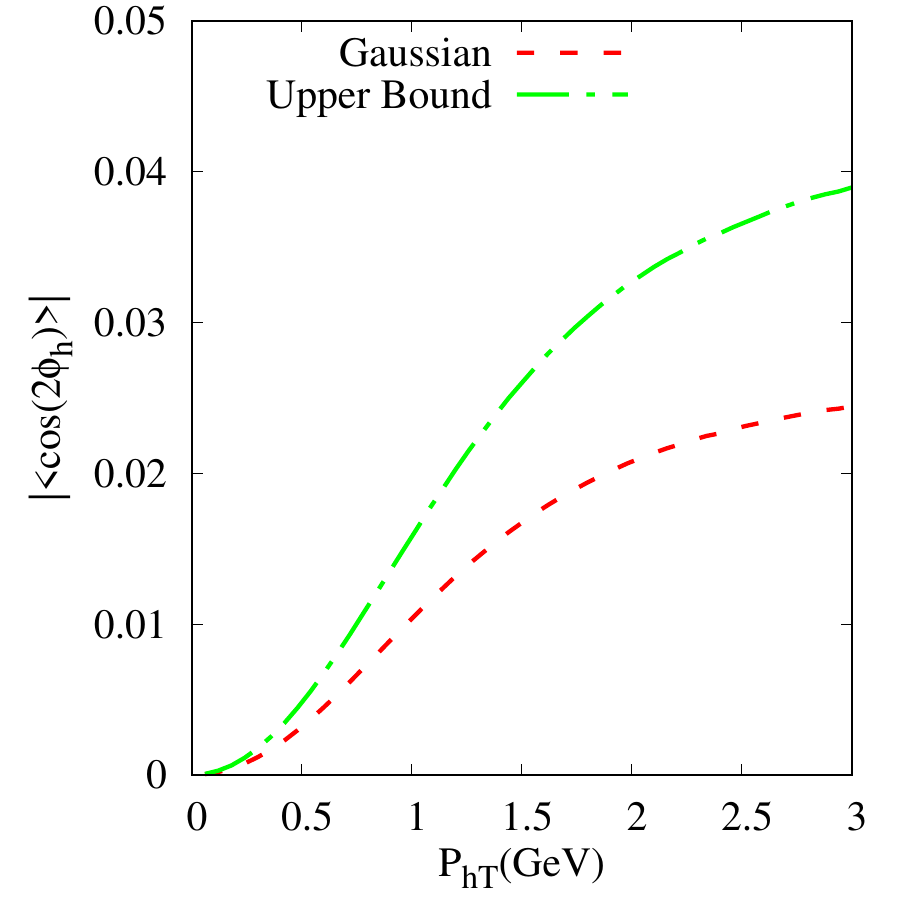}
		\hspace{0.1cm}
		\small{(b)}\includegraphics[width=7cm,height=6.5cm,clip]{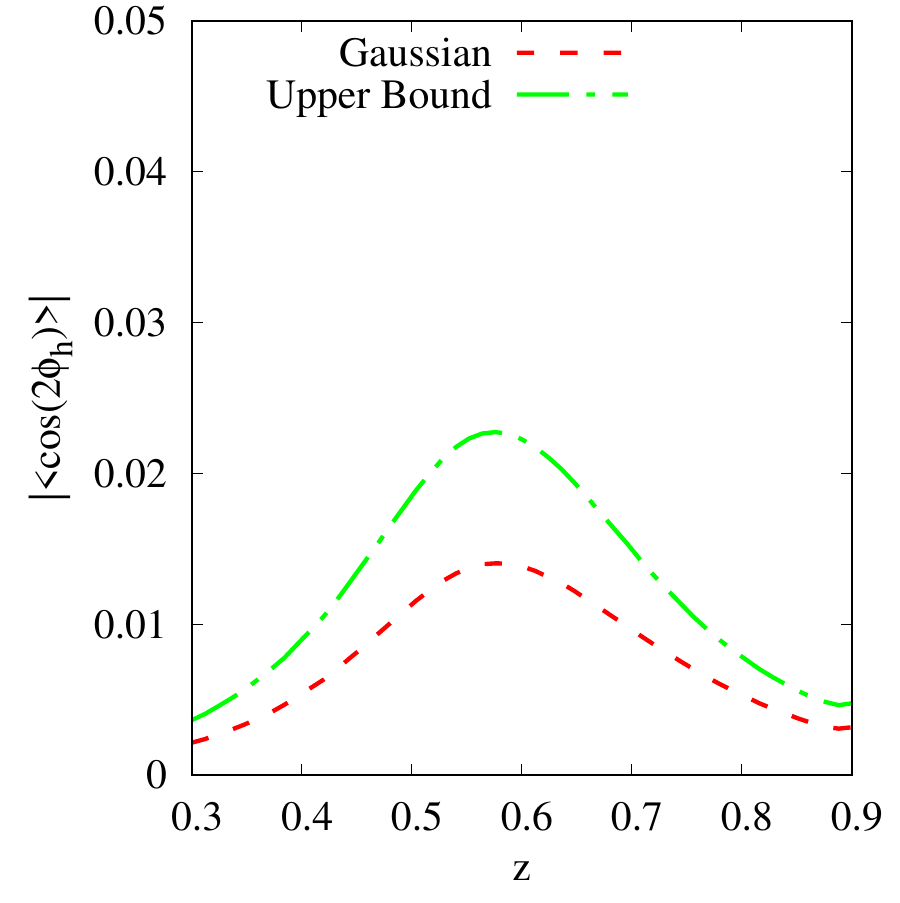}
	\end{minipage}
	\caption{\label{fig3}  $cos(2\phi_h)$ asymmetry  in $e+p\rightarrow e+ J/\psi +X$
		process as function of 
		(a) $P_{hT}$ (left panel) and  (b) $z$ (right panel) at $\sqrt{s}=150$ GeV (EIC) and $x_B=0.01$. The integration ranges are 
		$0<P_{hT}\leq3$ GeV, $0.3<z<0.9$ and $0.005<y<0.04$. For convention of lines see the legend in the plots.}
\label{figure3}
\end{figure}

\be
f_1^g(x,\textbf{k}_{\perp}^2)=\frac{S_{\perp}C_F}{\alpha_s\pi^3}\int dr\frac{J_0(k_{\perp}r)}{r}\left(1-e^{-\frac{r^2}{4}Q^2_{sg}(r)}\right)
\ee
\be
h_1^{\perp g}(x,\textbf{k}_{\perp}^2)=\frac{S_{\perp}C_F}{\alpha_s\pi^3}\frac{2M_p^2}{k_{\perp}^2}\int dr\frac{J_2(k_{\perp}r)}{r\log(\frac{1}{r^2\Lambda_{QCD}^2})}\left(1-e^{-\frac{r^2}{4}Q^2_{sg}(r)}\right)
\ee
where  $S_{\perp}$ is transverse size of the nucleus or nucleon. $Q_{sg}$ is the saturation scale, which in MV model, 
is defined as $Q_{sg}^2=\alpha_sN_c\mu_A\ln\frac{1}{r^2\Lambda_{QCD}^2}$ and $\mu_AS_{\perp}=\alpha_s2\pi A$, where $A=1$ for
 the proton. Following the approach of \cite{Bacchetta:2018ivt}, we have used a regularized version of the MV model in our calculation of the asymmetry.   The ratio of linearly polarized and unpolarized distribution in MV model can be given by

\be
\frac{\textbf{k}_{\perp}^2}{2M_p^2}\frac{h_1^{\perp g}(x,\textbf{k}_{\perp}^2)}{f_1^g(x,\textbf{k}_{\perp}^2)}=\frac{\int dr\frac{J_2(k_{\perp}r)}{r\log(\frac{1}{r^2\Lambda_{QCD}^2})}\left(1-e^{-\frac{r^2}{4}Q^2_{sg0}\log(\frac{1}{r^2\Lambda_{QCD}^2})}\right)}{\int dr\frac{J_0(k_{\perp}r)}{r}\left(1-e^{-\frac{r^2}{4}Q^2_{sg0}\log(\frac{1}{r^2\Lambda_{QCD}^2})}\right)}
\ee 
For $Q^2_{sg0}=(N_c/C_F)\times Q^2_{s0}$, where $Q^2_{s0}=0.35~{\mathrm{GeV}}^2$ at $x=0.01$ and $\Lambda_{QCD}=0.2~ {\mathrm{GeV}}$, the ratio is below $1$ for all $k_{\perp}$.  Below we give our numerical results.

\section{Numerical Results}

\begin{figure}[H]
	\begin{minipage}[c]{0.99\textwidth}
		\small{(a)}\includegraphics[width=7cm,height=6.5cm,clip]{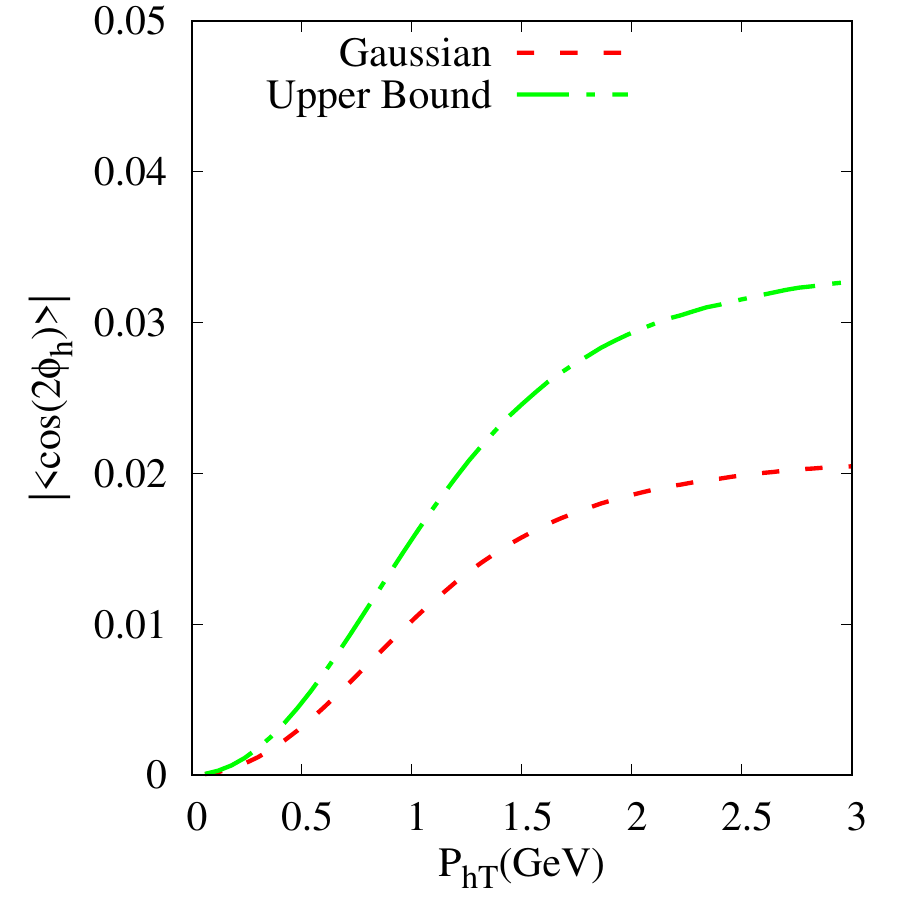}
		\hspace{0.1cm}
		\small{(b)}\includegraphics[width=7cm,height=6.5cm,clip]{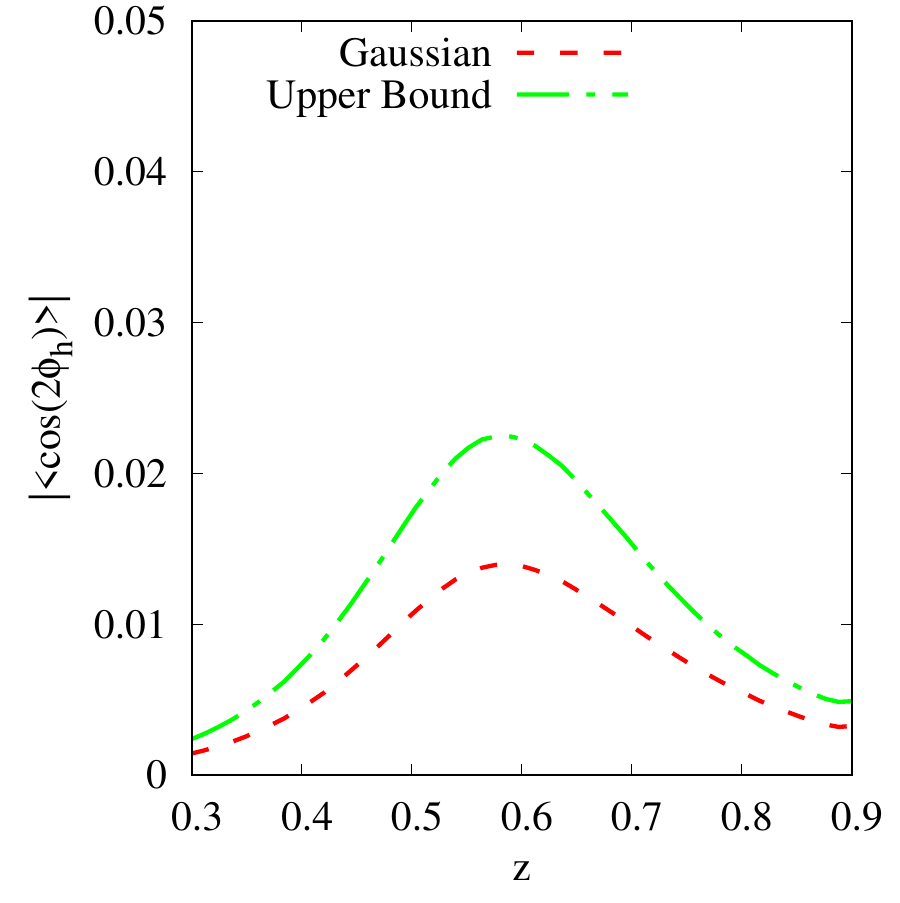}
	\end{minipage}
	\caption{\label{fig3}  $cos(2\phi_h)$ asymmetry  in $e+p\rightarrow e+ J/\psi +X$
		process as function of 
		(a) $P_{hT}$ (left panel) and  (b) $z$ (right panel) at $\sqrt{s}=190$ GeV (EIC) and $x_B=0.005$. The integration ranges are 
		$0<P_{hT}\leq3$ GeV, $0.3<z<0.9$ and $0.006<y<0.05$. For convention of lines see the legend in the plots.}
\label{figure4}
\end{figure} 

\begin{figure}[H]
			\begin{minipage}[c]{0.99\textwidth}
		\small{(a)}\includegraphics[width=7cm,height=6.5cm,clip]{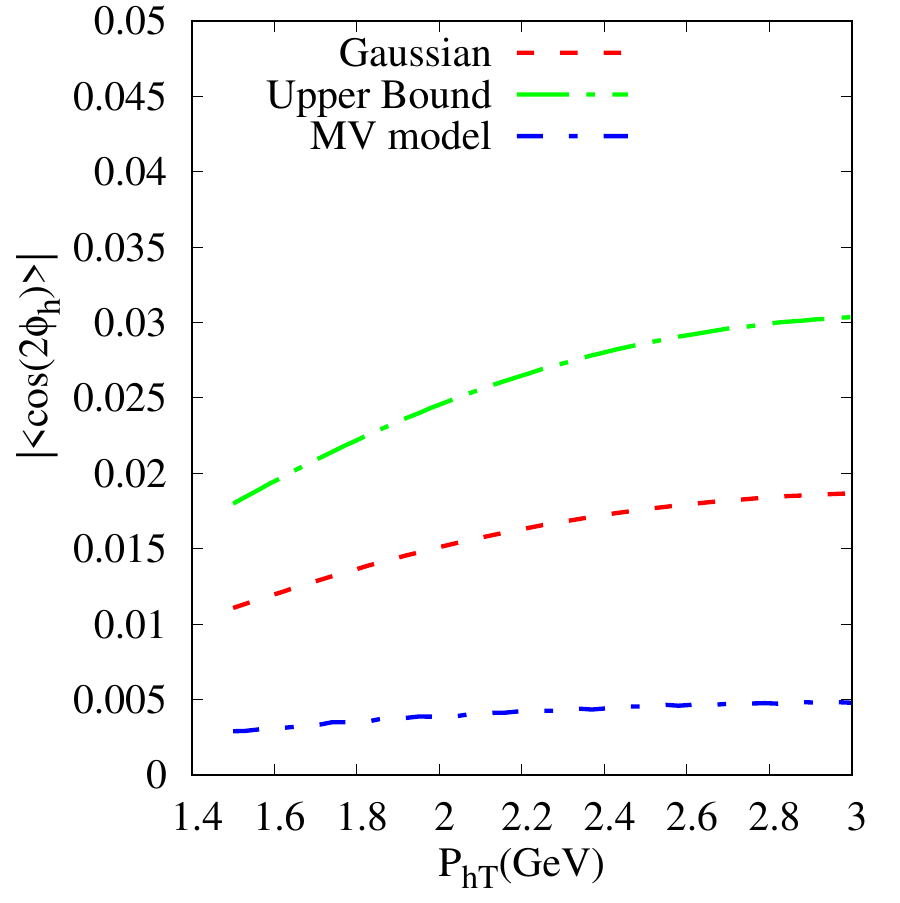}
		\hspace{0.1cm}
		\small{(b)}\includegraphics[width=7cm,height=6.5cm,clip]{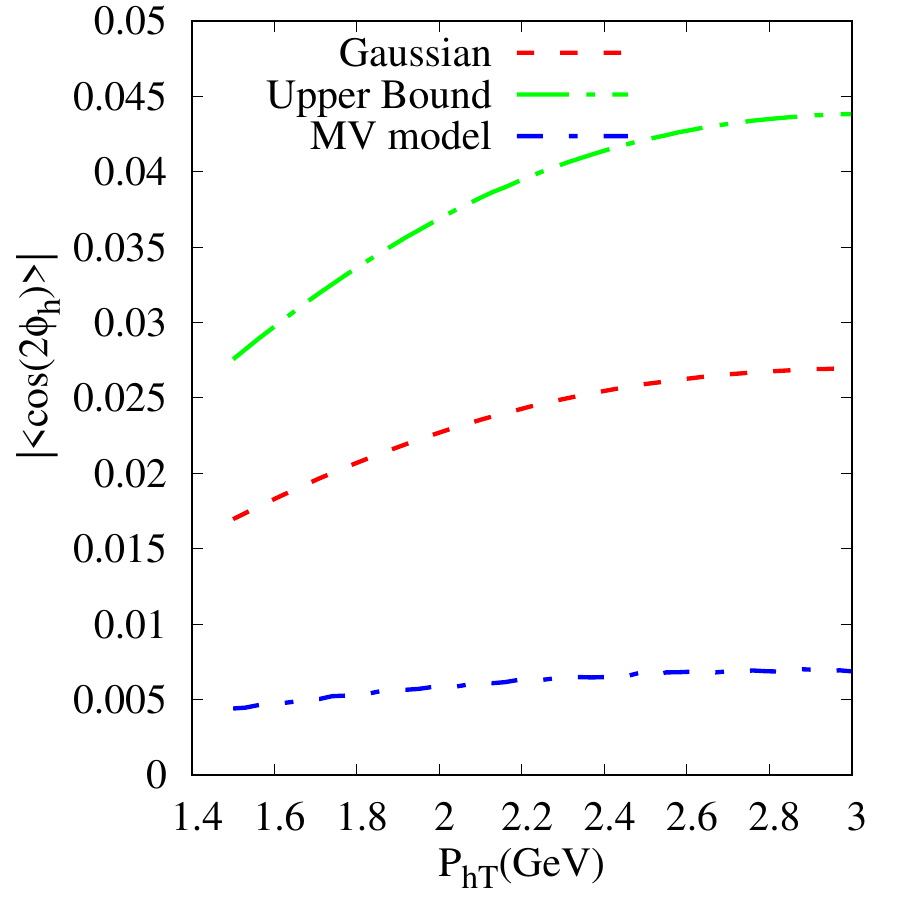}
			\end{minipage}
	\caption{\label{fig3}  $cos(2\phi_h)$ asymmetry  in $e+p\rightarrow e+ J/\psi +X$
		process as function of 
		$P_{hT}$ for (a) fixed $Q^2=9~{\mathrm{GeV}}^2$, $x=0.01$ and $z=0.5$ and (b)   fixed $Q^2=9~{\mathrm{GeV}}^2$, $x=0.01$ and $z=0.7$. In both the case integration  range on $y$ is $0.2<y<1$.  For convention of lines see the legend in the plots.}
\label{figure5}
\end{figure}

We have estimated the $cos(2\phi_h)$ asymmetry in $J/\psi$ production in the kinematics of EIC. MSTW2008 \cite{Martin:2009iq} is used for 
collinear PDFs. We have used the DGLAP evolution for the collinear pdfs. We have not included TMD evolution. As stated in the introduction, we have used  cuts on $z$, $0.3 < z < 0.9$. As we know, gluon initiated processes are enhanced at small $x$. In fact, small 
$x$ values will be accessed at EIC, and this kinematical region will be very important in determining the gluon TMDs including the linearly polarized gluon TMD. 
In this work, we have studied the $cos ~2 \phi$ asymmetry for EIC in the small $x$ region. It is to be noted that $x$ is related to the Bjorken variable $x_B$ through 
Eq. \ref{xb}. Smaller $x$ values also restrict $Q^2$ to be small, in this
work we took $Q^2$ to be of the same order and bounded by $M^2$ ($1< Q^2 < 9 ~\mathrm{GeV}^2$) , which is the mass of
$J/\psi$. For both the parametrizations used, the asymmetry is negative, which is consistent with the LO calculation \cite{Mukherjee:2016qxa}. 
 In the plots, we show the magnitude of the asymmetry.  CS LDMEs can be found for example in  \cite{Chao:2012iv}. As only one state contributes in the
 CS model, namely ${^3}{S}{_1}$ the asymmetry does not depend on the specific set of LDME. This is different from the CO model, where even at LO,
 contribution comes from several states \cite{Mukherjee:2016qxa}, and the result depends on the choice of LDMEs. However, the unpolarized cross section will depend on the choice of LDMEs in both the models. In our previous work \cite{Mukherjee:2016qxa}, we compared with three set of LDMEs where one sets of LDMEs giving the unpolarized cross section that matches more with the experimental data than the other sets. Figs. \ref{figure2} , \ref{figure3} and \ref{figure4}  show the  upper bound of the asymmetry  as well as an estimate using the Gaussian model, 
at $\sqrt{s}=45$~GeV, $150$~GeV and $190$~GeV respectively, as a function of $P_{hT}$ and $z$ .
 Corresponding $x_B$ values are $x_B=0.01,~0.01$ and $0.005$ respectively;  ranges of $y$ integration are $0.05<y<0.4,~0.005<y<0.04$ and 
$0.006<y<0.05$ respectively.  $y$ is constrained by the choice of $Q^2$ and $x_B$.  The transverse momentum $P_{h\perp}$ of $J/\psi$ is 
taken in range $0<P_{h\perp}<3$~ GeV. Energy fraction $z$ is in the range $0.3<z<0.9$ for all these plots. 
 The upper bound of the asymmetry increases with increase of  $\sqrt{s}$ for the same $x_B$, it reaches maximum near $P_{hT} 
\approx 3$~ GeV, the maximum is about $4\%$ for $\sqrt{s}=150$~GeV.  However, for smaller $x_B$, asymmetry decreases.
 The asymmetry reaches a peak near $z=0.6$ for the kinematical cuts chosen. The qualitative behavior of the asymmetry remains the 
same for all $\sqrt{s}$. The Gaussian model gives smaller asymmetry.  Fig. \ref{figure5} shows a comparison of the upper bound of 
the asymmetry with that calculated in Gaussian model as well as MV model, as a function of $P_{hT}$, for two different values of $z$, 
(a) $z=0.5$ and (b)  $z= 0.7$. For both these plots we have taken  fixed value of $x=0.01$, $Q^2=9~{\mathrm{GeV}}^2$ and $0.2 < y <1$. For (a)
   $\sqrt{s}$ is in the range $61$ to $181$~ GeV, and for (b)  $\sqrt{s}$ is in the range $58$ to  $182$~ GeV. Asymmetry in the MV model is 
smaller compared to the Gaussian model, and both  lie below the upper bound.  The asymmetry is higher for higher values of $z$.

\section{conclusion}

In this work, we have calculated the $cos~2 \phi$ asymmetry in electroproduction of $J/\psi$ at EIC, that probes the linearly polarized gluon distribution in the unpolarized proton. 
We calculated the asymmetry in the kinematical region $z<1$, where the NLO subprocess $\gamma^* +g \rightarrow J/\psi +g$ gives the leading contribution. The gluon TMDs probed
in this process are of Weizs\"acker-Williams (WW) type. As gluon distributions pay an important role in the small $x$ region,  we investigate the asymmetry in the small $x$ kinematical 
region, using a Gaussian parametrization of the TMDs as well as in McLerran-Venugopalan  model. We also show the upper bound of the asymmetry saturating the inequality for the 
linearly polarized gluon distribution. At EIC, low values of $x$ also restrict the $Q^2$ (virtuality of the photon) values.  We have calculated the $J/\psi$ production amplitude 
in NRQCD based color singlet  (CS) approach. The asymmetry in the kinematical region considered is small but sizable. The magnitude of the asymmetry may depend on the production 
mechanism of the quarkonium. As shown in \cite{Mukherjee:2016qxa}, CS mechanism underestimates the $J/\psi$ production at HERA, and both CS and CO contributions are needed to describe 
the data. In CO formalism contribution will come from several LDMEs  in the final state, which may enhance the asymmetry. We plan to see the effect of the CO mechanism on the asymmetry
 in a future work. Another interesting study would be the effect of small-$x$ evolution on the asymmetry.  In any case, the $cos~2 \phi$ asymmetry in $J/\psi$ production at the EIC  will 
be an important tool to gain information on the WW type linearly polarized gluon distribution. 

\section{Acknowledgement}
We thank P. J. Mulders, E. Petreska  and D. Boer for discussion. Part of the work of AM was done at  NIKHEF, Amsterdam, the visit was supported by the European Research Council under the "Ideas" program QWORK  (contract 320389).

\section{Appendix}

All the amplitude squares and the coefficients are integrated over $\phi$, where $\phi$ and $\phi_h$ are the azimuthal angle of initial gluon and $J/\psi$ respectively.
\be
M'_iM'_j=\int d\phi M_iM_j
\ee

\begin{equation}
\begin{split}
M'_1M'_1=&\big\{f_1^g\times 128 \pi  M^4 \big\{M^4 (z-1)^3+M^2 \big({P_{h\perp}}^2(z-1) (8 (z-1) z+3)\\
&+s y \big({x_B} \big(z \big((1-6y (y+2)) z^2\\
&+2 y (y+4) z-2 y+z+1\big)-1\big)-x(z-1)^3\big)\big)\\
&+2 {P_{h\perp}} \sqrt{{x_B}}\sqrt{1-y} z \sqrt{s y} \cos ({\phi_h}) \big(M^2 (z(4 z-5)+2)\\
&+{P_{h\perp}}^2 (z (24 z-17)+4)-2 s x y (z (4 z-3)+1)\big)\\
&+{P_{h\perp}}^4 (z (z (11z-23)+13)-3)+{P_{h\perp}}^2 s y \big(x (z (-2 (z-5)z-9)+3)\\
&+{x_B} \left(z \left(-3 (y (5 y+16)-6) z^2+2 (y(y+14)-5) z-6 y+7\right)-3\right)\big)\\
&+s^2 x {x_B}y^2 z (z ((y-2) y (2 z-1)+2 (8 z-7))+4)\big\}\\
%
&+\frac{k_{\perp}^2h_1^{\perp g}}{M_p^2}\times 64 \pi  M^4 \big\{M^4\left(-(z-1)^3\right)+M^2\big({P_{h\perp}}^2 (z (-4 (z-2) z-7)+3)\\
&+s y \big(x(z-1)^3+{x_B} z \big((y (7 y+18)-7) z^2-2 y (2 y+5)z\\
&+2y+z-1\big)+{x_B}\big)\big)+2 {P_{h\perp}}\sqrt{{x_B}} z \big({P_{h\perp}} s \sqrt{{x_B}} (y-1)y (z (5 z-4)+2)\\
&\times\cos (2\phi_h)-\sqrt{1-y} \sqrt{sy} \cos (\phi_h) \big(M^2 (1-3 z)^2+{P_{h\perp}}^2(z (13 z-12)+5)\\
&+s x y (z (6z-5)+2)\big)\big)+{P_{h\perp}}^4 (z ((7-3 z)z-6)+3)\\
&+{P_{h\perp}}^2 s y \big({x_B} \big(z \big((3 y(3 y+8)-22) z^2-2 (y (y+13)-8) z\\
&+10 y-9\big)+3\big)-x(z (2 z (z+1)-5)+3)\big)\\
&-s^2 x y^2 z \big(x \left(6z^2-4 z+1\right)+{x_B} z \big((y-2) y (4 z-3)\\
&+26 z-20\big)-2{x_B} (y-3)\big)\big\}\big\}/\big\{x y^2 (z-1)^2 z \left(M^2+sy ({x_B}-x)\right)^2\\
&\times\left(M^2+{P_{h\perp}}^2-s {x_B}y (z-2) z\right)^2\big\}\\
\end{split}
\end{equation}

\begin{equation}
\begin{split}
M'_1M'_2=&\big\{f_1^g\times 64 \pi  M^6 \big\{M^2 (z-1) (2 z-3)+{P_{h\perp}}^2 (2 z (4 z-9)+9)\\
&-2 {P_{h\perp}} \sqrt{{x_B}} \sqrt{1-y} (z-1) z (3 z-8) \sqrt{s y} \cos ({\phi_h})\\
&-s y (z-1) \big(x
(z+1) (2 z-3)+{x_B} z \big(3 (y (y+6)-6) z^2\\
&-6 (y(y+4)-3) z+4 y+3\big)+2 {x_B}y+{x_B}\big)\big\}\\
&-\frac{k_{\perp}^2h_1^{\perp g}}{M_p^2}\times 32 \pi  M^6 \big\{M^2 (z-1) (2 z-3)+{P_{h\perp}}^2 (2 z (3 z-7)+7)\\
&-2 {P_{h\perp}} \sqrt{{x_B}} \sqrt{1-y} (z-1) (2 z-3) (3 z+1) \sqrt{s y} \cos ({\phi_h})\\
&-s y (z-1) \big(x (3 z-5)+{x_B} z \big((y (5 y+6)-6) z^2-2 y (5 y+7) z\\
&+2 y (y+2)+6 z+5\big)+{x_B} (4 y-1)\big)\big\}\big\}/\big\{s x^2 y^3 (z-1) z^3\\
&\times \left(M^2+s y ({x_B}-x)\right)^2 \left(M^2+{P_{h\perp}}^2-s {x_B} y (z-2) z\right)\big\}
\end{split}
\end{equation}

\begin{equation}
\begin{split}
M'_1M'_3=&\big\{-f_1^{g}\times 64 \pi  M^8 \big\{2 M^2 (z-1)^2 (3 z-5)+4 {P_{h\perp}}^2 (z-1) (z (6 z-13)+8)\\
&-2 {P_{h\perp}} \sqrt{{x_B}} \sqrt{1-y} (z-1)^2 (z (14 z-25)+2) \sqrt{s y} \cos ({\phi_h})\\
&+s y \big(-x (3 z-10) (z-1)^2+{x_B} z \big(y^2 (z (z (31 z-57)+43)-9)\\
&+2 y (z (z (31  z-59)+53)-23)-2 z (5 z (5 z-8)+39)+56\big)\\
&-2 {x_B} ((y-4)  y+10)\big)\big\}\\
&+\frac{k_{\perp}^2h_1^{\perp g}}{M_p^2}\times 32 \pi  M^8 \big\{2 M^2 (z-1)^2 (7 z-9)+2 {P_{h\perp}}^2 (z-1) (z (14 z-41)+29)\\
&-4 {P_{h\perp}} \sqrt{{x_B}} \sqrt{1-y} (z-1)^2 (7 z (2 z-3)+3) \sqrt{s y} \cos ({\phi_h})\\
&+s y \big({x_B} \big(y^2 (z (z (z (55 z-79)+45)-3)-4)\\
&+2 y (z (z (z (59 z-96)+76)-30)+5)\\
&+2 z (z  ((51-43 z) z-44)+37)-30\big)\\
&-x (z-1)^2 (3 z (4 z+1)-14)\big)\big\}\big\}/\big\{s x^2 y^3 (z-1)^2 z^2 \left(M^2+s y ({x_B}-x)\right)^2\\
&\times \left(M^2+{P_{h\perp}}^2-s {x_B} y (z-2) z\right)^2\big\}
\end{split}
\end{equation}

and the coefficients in final expression of cross section, Eq.~(\ref{csf}), are as follows: 

\begin{equation}
\begin{split}
A_0=& 64 \pi  M^4 \big\{\big\{M^2 (z-1) \left(M^2+{P_{h\perp}}^2-s{x_B} y (z-2) z\right) \big(M^2 (z-1) (2 z-3)\\
&+{P_{h\perp}}^2(2 z (4 z-9)+9)
-s y (z-1) \big(x (z+1) (2 z-3)\\
&+{x_B} z\left(3 (y (y+6)-6) z^2-6 (y (y+4)-3) z+4 y+3\right)+2 {x_B}y+{x_B}\big)\big)\\
&+M^4 z \big(-2 M^2 (z-1)^2(3 z-5)-4 {P_{h\perp}}^2 (z-1) (z (6 z-13)+8)\\
&+s y \big(x (3 z-10)(z-1)^2+{x_B} z \big((50-31 y (y+2)) z^3\\
&+(y (57 y+118)-80)z^2-y (43 y+106) z\\
&+y (9 y+46)+78 z-56\big)+2 {x_B} ((y-4)y+10)\big)\big)\big\}/s\\
&+2 x y z^2 \big(M^4 (z-1)^3+M^2\big({P_{h\perp}}^2 (z-1) (8 (z-1) z+3)\\
&+s y \big({x_B}\left(z \left((1-6 y (y+2)) z^2+2 y (y+4) z-2y+z+1\right)-1\right)\\
&-x (z-1)^3\big)\big)+{P_{h\perp}}^4 (z (z(11 z-23)+13)-3)\\
&+{P_{h\perp}}^2 s y \big(x (z (-2 (z-5)z-9)+3)+{x_B} \big(z \big(-3 (y (5 y+16)-6) z^2\\
&+2 (y(y+14)-5) z-6 y+7\big)-3\big)\big)\\
&+s^2 x {x_B} y^2 z(z ((y-2) y (2 z-1)+2 (8 z-7))+4)\big)\big\}/\big\{x^2 y^3 (z-1)^2
z^3 \\
&\times\left(M^2+s y ({x_B}-x)\right)^2 \left(M^2+{P_{h\perp}}^2-s
{x_B} y (z-2) z\right)^2\big\}
\end{split}
\end{equation}

\begin{equation}
\begin{split}
A_1=&-128 \pi  M^4 {P_{h\perp}} \sqrt{{x_B}} \sqrt{1-y} \big\{-2M^4 (z-1)^2 (7 (z-2) z+5)\\
&+M^2 \big({P_{h\perp}}^2 (z-1)^2 (3z-8)-s y z \big(2 x z (z (4 z-5)+2)\\
&+{x_B} (z-2) (3 z-8)(z-1)^2\big)\big)+2 s x y z^2 \big({P_{h\perp}}^2 ((17-24 z)z-4)\\
&+2 s x y (z (4 z-3)+1)\big)\big\}/\big\{x^2 y^2 (z-1)^2 z^2\sqrt{s y} \left(M^2+s y ({x_B}-x)\right)^2\\
&\times\left(M^2+{P_{h\perp}}^2-s {x_B} y (z-2) z\right)^2\big\}
\end{split}
\end{equation}

\begin{equation}
\begin{split}
B_0=& 32 \pi  M^4 \big\{\big\{-M^2 (z-1) \big(M^2+{P_{h\perp}}^2-s{x_B} y (z-2) z\big) \big(M^2 (z-1) (2 z-3)\\
&+{P_{h\perp}}^2(2 z (3 z-7)+7)-s y (z-1) \big(x (3 z-5)+{x_B} z \big((y(5 y+6)-6) z^2\\
&-2 y (5 y+7) z+2 y (y+2)+6 z+5\big)+{x_B} (4y-1)\big)\big)\\
&+M^4 z \big(2 M^2 (z-1)^2 (7 z-9)+2{P_{h\perp}}^2 (z-1) (z (14 z-41)+29)\\
&+s y \big({x_B} \big(y^2(z (z (z (55 z-79)+45)-3)-4)\\
&+2 y (z (z (z (59 z-96)+76)-30)+5)\\
&+2z (z ((51-43 z) z-44)+37)-30\big)\\
&-x (z-1)^2 (3 z (4z+1)-14)\big)\big)\big\}/{s}\\
&-2 x y z^2 \big(M^4 (z-1)^3+M^2\big({P_{h\perp}}^2 (z-1) (4 (z-1) z+3)\\
&-s y \big(x(z-1)^3+{x_B} z \big((y (7 y+18)-7) z^2\\
&-2 y (2 y+5) z+2y+z-1\big)+{x_B}\big)\big)\\
&+{P_{h\perp}}^4 (z (z (3z-7)+6)-3)+{P_{h\perp}}^2 s y \big(x (z (2 z (z+1)-5)+3)\\
&+{x_B}\big(z \big((22-3 y (3 y+8)) z^2+2 (y (y+13)-8) z\\
&-10y+9\big)-3\big)\big)+s^2 x y^2 z \big(x \big(6 z^2-4z+1\big)\\
&+{x_B} z ((y-2) y (4 z-3)+26 z-20)-2 {x_B}(y-3)\big)\big)\big\}/\big\{x^2 y^3 (z-1)^2 z^3\\
&\times \left(M^2+s y({x_B}-x)\right)^2 \left(M^2+{P_{h\perp}}^2-s {x_B} y (z-2)z\right)^2\big\}
\end{split}
\end{equation}

\begin{equation}
\begin{split}
B_1=&- 64 \pi  M^4 {P_{h\perp}} \sqrt{{x_B}} \sqrt{1-y} \big\{M^4(z-1)^2 (z (4 z (7 z-12)+13)+3)\\
&+M^2 \big(s y z \left(2 x (1-3z)^2 z^2+{x_B} (z-2) (2 z-3) (3 z+1)
(z-1)^2\right)\\
&-{P_{h\perp}}^2 (z-1)^2 (2 z-3) (3 z+1)\big)+2 s xy z^3 \big({P_{h\perp}}^2 (z (13 z-12)+5)\\
&+s x y (z (6
z-5)+2)\big)\big\}/\big\{x^2 y^2 (z-1)^2 z^3 \sqrt{s y}
\left(M^2+s y ({x_B}-x)\right)^2 \\
&\times\left(M^2+{P_{h\perp}}^2-s
{x_B} y (z-2) z\right)^2\big\}
\end{split}
\end{equation}

\begin{equation}
\begin{split}
B_2=&\big\{128 \pi  M^4 {P_{h\perp}}^2 s {x_B} (y-1) (z (5 z-4)+2)\big\}/\big\{x y
(z-1)^2 \\
&\times\left(M^2+s y ({x_B}-x)\right)^2\left(M^2+{P_{h\perp}}^2-s {x_B} y (z-2) z\right)^2\big\}
\end{split}
\end{equation}

\bibliographystyle{apsrev}
\bibliography{refer}
\end{document}